\documentclass[12pt]{iopart}

\expandafter\let\csname equation*\endcsname\relax
\expandafter\let\csname endequation*\endcsname\relax

\usepackage{hyperref}
\usepackage{amsmath}
\usepackage[left=1.5in,right=1in,top=1in,bottom=1in,includefoot]{geometry}
\usepackage{bm}
\usepackage{graphics}
\usepackage{graphicx}
\usepackage{epsfig}
\usepackage{color}
\usepackage{subfig}
\usepackage{rotating}
\usepackage{latexsym}
\usepackage{textcomp}
\usepackage{appendix}
\usepackage{multirow}
\usepackage{url}
\usepackage[acronym,shortcuts]{glossaries}
\usepackage{mathtools}
\usepackage{nicefrac}
\usepackage{fullpage}
\usepackage{verbatim}
\usepackage{setspace}

\newglossary[slg]{symbols}{sym}{sbl}{List of Terminology}
\makeglossaries
\newacronym{LIGO}{LIGO}{Laser Interferometer Gravitational-Wave Observatory}
\newacronym{aLIGO}{aLIGO}{Advanced LIGO}
\newacronym{LHO}{LHO}{LIGO Hanford Observatory}
\newacronym{LLO}{LLO}{LIGO Livingston Observatory}
\newacronym{LSC}{LSC}{LIGO Scientific Collaboration}
\newacronym{CBC}{CBC}{compact binary coalescence}
\newacronym{GW}{GW}{gravitational wave}
\newacronym{DQ}{DQ}{data quality}
\newacronym{SNR}{SNR}{signal-to-noise ratio}
\newacronym{S4}{S4}{LIGO Science Run 4}
\newacronym{S5}{S5}{LIGO Science Run 5}
\newacronym{S6}{S6}{LIGO Science Run 6}
\newacronym{HEPI}{HEPI}{hydraulic external pre-isolator}
\newacronym{KW}{KW}{Kleine-Welle}
\newacronym{HVeto}{HVeto}{HierarchichalVeto}
\newacronym{PEM}{PEM}{physical environment monitor}
\newacronym{DetChar}{DetChar}{Detector Characterisation}
\newacronym{LVEA}{LVEA}{Large Vacuum Equipment Area}
\newacronym{OMC}{OMC}{Output Mode Cleaner}
\newacronym{USDOE}{USDOE}{United States Department of Energy}
\newacronym[plural={glitches}, shortplural={glitches}]{glitch}{glitch}{transient
 noise event}
\newacronym{PN}{PN}{post-Newtonian}
\newacronym{PSD}{PSD}{power spectral density}
\newacronym{DARM}{DARM}{differential arm}
\newacronym{CARM}{CARM}{common arm}
\newacronym{MICH}{MICH}{Michelson}\newacronym{GR}{GR}{general relativity}
\newacronym{VSR4}{VSR4}{Virgo Science Run 4}\newacronym{GRB}{GRB}{gamma ray burst}
\newacronym{GWB}{GWB}{gravitational-wave burst}\newacronym{GEO-HF}{GEO-HF}{GEO High Frequency}
\newacronym{FFT}{FFT}{fast Fourier transform}
\newglossaryentry{veto}{name={veto}, description={time segment indicating poor \gls{DQ} to be removed from an analysis}, descriptionplural={time segments indicating poor \gls{DQ} which are removed from an analysis}, plural={vetoes}}
\newglossaryentry{trigger}{name={trigger}, description={event produced by a \ac{GW} search algorithm}, descriptionplural={events produced by a \ac{GW} search algorithm}}
\newglossaryentry{upconversion}{name={upconversion}, description={non-linear coupling of low-frequency noise into higher frequency/broadband noise}}
\newglossaryentry{light dip}{name={light dip}, description={drop in the power stored in a detector arm cavity}, descriptionplural={drops in the power stored in a detector arm cavity}}
\newglossaryentry{efficiency}{name={efficiency}, description={the fractional number of GW triggers removed by a veto}}
\newglossaryentry{deadtime}{name={deadtime}, description={the fractional amount of analysis time that has been vetoed}}
\newglossaryentry{SeisVeto}{name={SeisVeto}, description={the \ac{LIGO} seismic veto developed using targeted veto methods in \ac{S6} \cite{Macleod:2011up}}}
\glsdisablehyper

\makeatletter

\newcommand{\Rmnum}[1]{\expandafter\@slowromancap\romannumeral #1@}
\makeatother

\let\origdescription\description
\renewenvironment{description}{
  \setlength{\leftmargini}{0em}
  \origdescription
  \setlength{\itemindent}{0em}
  \setlength{\labelsep}{\textwidth}
}
{\endlist}

\begin{document}

\title{Reducing the effect of seismic noise in LIGO searches by targeted veto generation}
 
\author{
D.~M.~Macleod$^{1}$,
S.~Fairhurst$^{1}$,
B.~Hughey$^{2}$,
A.~P.~Lundgren$^{3,4}$,
L.~Pekowsky$^{4}$,
J.~Rollins$^{5}$,
J.~R.~Smith$^{4,6}$}
\address{$^{1}$Cardiff University, Cardiff, CF24 3AA, United Kingdom }
\address{$^{2}$University of Wisconsin--Milwaukee, Milwaukee, WI  53201, USA }
\address{$^{3}$The Pennsylvania State University, University Park, PA  16802, USA }
\address{$^{4}$Syracuse University, Syracuse, NY  13244, USA }
\address{$^{5}$LIGO - California Institute of Technology, Pasadena, CA  91125, USA }
\address{$^{6}$California State University Fullerton, Fullerton CA 92831, USA}
\ead{duncan.macleod@astro.cf.ac.uk}
 
\begin{abstract}
One of the major obstacles to the detection and study of \aclp{GW} using ground-based laser interferometers is the effect of seismic noise on instrument sensitivity.
Environmental disturbances cause motion of the interferometer optics, coupling as noise in the \acl{GW} data output whose magnitude can be much greater than that of an astrophysical signal.
We present an improved method of identifying times of high seismic noise coupling by tuning a \acl{GWB} detection algorithm to the low-frequency signature of these events and testing for coincidence with a low-latency \acl{CBC} detection algorithm.
This method has proven highly effective in removing transients of seismic origin, with 60\% of all \acl{CBC} candidate events correlated with seismic noise in just 6\% of analysis time.

\end{abstract}
\maketitle

\glsresetall

\section{Introduction}\label{sec:intro}
\noindent
\noindent The \ac{LIGO} \cite{Abbott:2007kv} is designed to detect and study \acp{GW} of astrophysical origin.
During \ac{S6} the project operated two kilometer-scale, power-recycled, Michelson interferometers in the United States, at Hanford, WA and  Livingston, LA.
Together with the French-Italian Virgo \cite{Acernese:2008zza} detector they were involved in the search for \ac{GW} signals from many signals, including the coalescence of compact binary systems \cite{Abadie:2010yb}, and unmodelled burst events \cite{Abadie:2010mt}.

The output of each \ac{LIGO} detector is a single data stream that in general contains some combination of a \ac{GW} signal and detector noise.
\Acp{glitch} can mask or mimic astrophysical signals, thus limiting the sensitivity of any search that can be performed over these data \cite{Blackburn:2008ah, Christensen:2010zz, Abadie:2011dc}.
In the searches for short-duration \ac{GW} signals the noise background is dominated by \acp{glitch}, requiring intense effort from analysis groups and detector scientists to understand the physical origins and eliminate them.
Throughout the lifetime of \ac{LIGO} up to and including \ac{S6}, search sensitivity has been improved by careful use of \glspl{veto}. \Glspl{veto} allow analysts to tune and operate search pipelines using a subset of cleaner data, increasing the chance of extracting a signal from the noise \cite{Slutsky:2010ff}.

The detrimental effect of seismic noise has been known to be a key limiting factor to the sensitivity of \ac{GW} detectors at low frequencies (below a hundred Hz \cite{Daw:2004qd}).
However it is also a common cause of \acp{glitch} at higher frequencies due to non-linear coupling of low-frequency seismic noise into the gravitational wave readout.
Previous methods to generate \gls{veto} segments for times of high seismic noise have proven ineffective.
In this paper, we introduce a new method of constructing vetoes for the specific case of seismic noise that has proven highly effective when used in the latest searches for transient \ac{GW} signals. We find a large statistical correlation between \glspl{trigger} from the low-latency \ac{CBC} search and seismic noise, vetoing 60\% of all \glspl{trigger} in 6\% of time for H1 and 6\% of \glspl{trigger}  in 0.6\% of time for L1.  

The paper is set out as follows.
Section \ref{sec:seismic} describes the seismic environment at each of the \ac{LIGO} sites, and the effect it has on detector sensitivity.
In section \ref{sec:existing} we outline existing \gls{veto} methods used in \acs{S5} and \acs{S6}.
In section \ref{sec:method} we describe our new method for identifying and vetoing noise in seismometer data.
In section \ref{sec:results} we present the results in terms of veto efficiency and deadtime.
Finally, section \ref{sec:summary} presents a brief discussion of implications and further applications of the method.

\section{Seismic noise in LIGO}\label{sec:seismic}
\noindent
\subsection{LIGO seismic environment}
\begin{figure}
  \centering
  \includegraphics[width=0.9\linewidth]{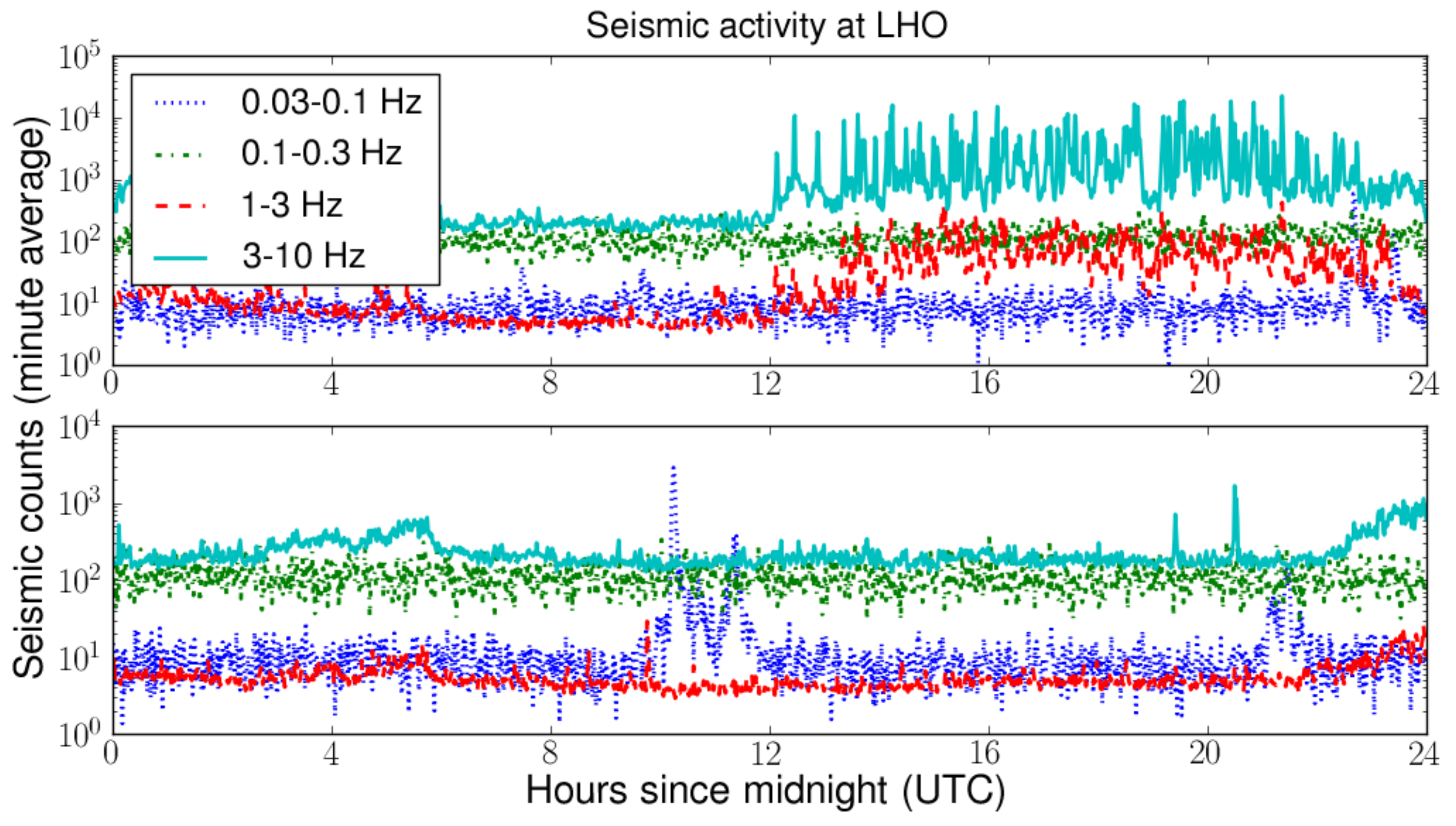}
  \caption{\label{fig:seisdiff}
Examples of the difference between seismic-induced acceleration incident at the LHO site during a full 24-hour span on a weekday (top) and a weekend (bottom), as measured by a seismometer. The higher frequency bands show elevated noise on a weekday between 12:00--24:00 UTC (05:00-17:00 local time) from working-day traffic.}
\end{figure}

\begin{figure}[htp]
  \centering
  \includegraphics[width=0.8\linewidth]{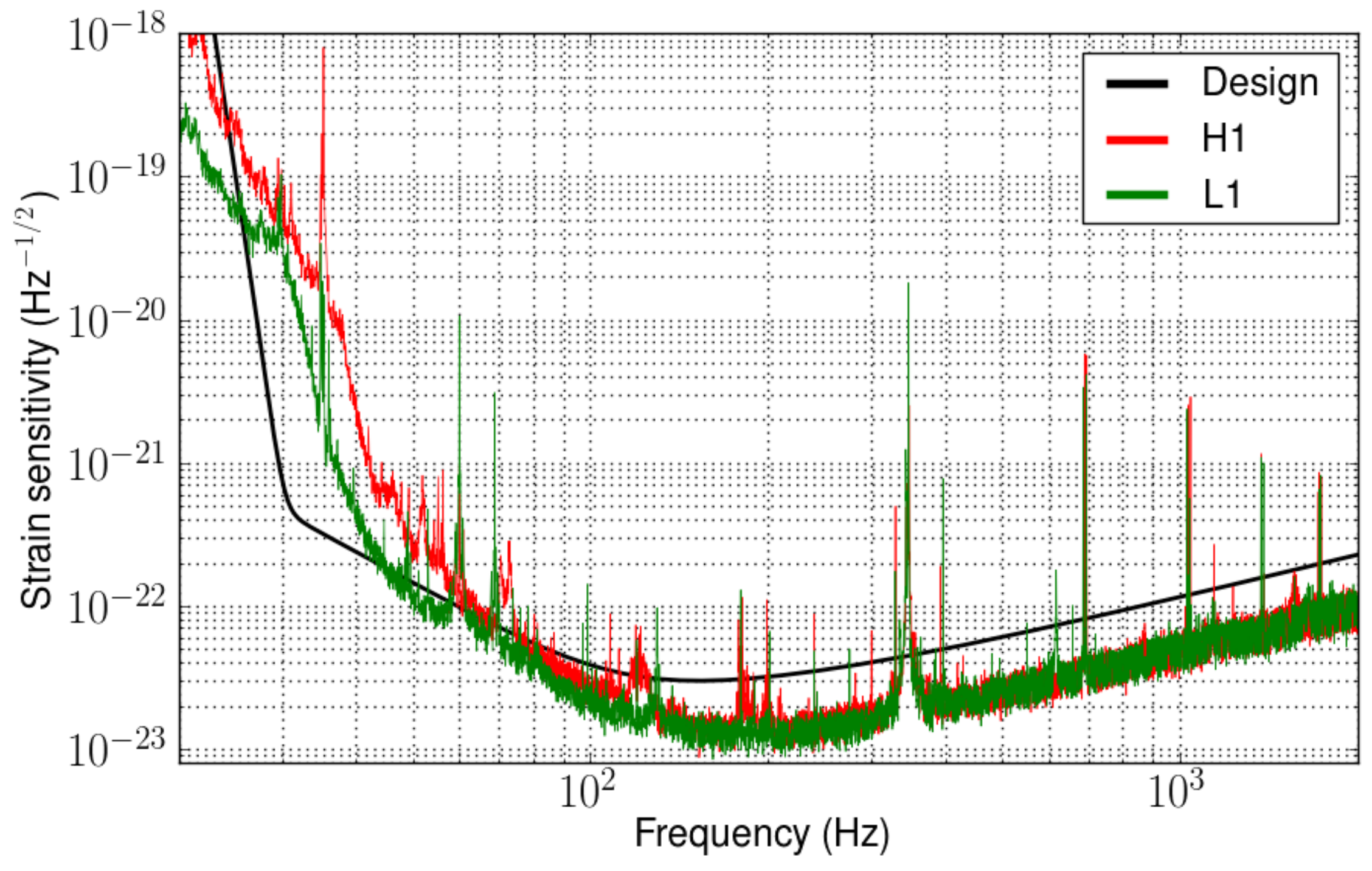}
  \caption{\label{fig:ligonoise}
Strain sensitivity of the \ac{LIGO} interferometers during \ac{S6}.
The seismic background (from direct coupling only) is the limiting noise source below $\sim50$\,Hz.}
\end{figure}

The two \ac{LIGO} sites were chosen to be far from urbanised areas, thus reducing the incident seismic noise, whilst their separation provides a long baseline helpful in sky-localisation of astrophysical signals \cite{Fairhurst:2009tc}.
The various types of seismic noise to which they are subject, as characterised by their source, can be separated into the four frequency bands given in Table \ref{table:seismicbands}, and the variablility of noise during evenings and weekends relative to standard working hours is shown in Figure \ref{fig:seisdiff}.
The strain sensitivity of the two \ac{LIGO} detectors is shown in Figure \ref{fig:ligonoise}, with seismic noise the limiting factor below 50\,Hz.
\begin{table}
\begin{center}  \begin{tabular}{c|c|l}
    \hline
    \hline
    Frequency (Hz) & Distance (km) & Source\\
    \hline
    \hline
    $0.01-1$ & $10^3$    & \begin{tabular}{l}
                             Distant earthquakes\\
                             Microseism\\
                           \end{tabular}\\
    \hline
    $1-3$    & $10^1$    & \begin{tabular}{l}
                             Far anthropogenic noise\\
                             Close earthquakes\\
                             Wind\\
                           \end{tabular}\\
    \hline
    $3-10$   & $10^0$    & \begin{tabular}{l}
                             Anthropogenic noise\\
                             Wind\\
                           \end{tabular}\\
    \hline
    $10-100$ & $10^{-1}$ & \begin{tabular}{l}
                             Close anthropogenic noise\\
                           \end{tabular}\\
  \hline\hline
  \end{tabular}
\caption{Description of the main seismic frequency bands and their sources}\label{table:seismicbands}
\end{center}
\end{table}

\ac{LHO} is located 15\,km from the \ac{USDOE} Hanford Site, in which several working areas include use of heavy earth-moving machinery, and the Tri-Cities area begins roughly 20\,km away, both contributing heavily to the amount of anthropogenic seismic noise incident on the detector.
The site is also subject to high winds up to 40\,m/s, causing motion of the buildings and the concrete slabs supporting the instruments.
These relatively local sources generate noise in the higher bands in Table \ref{table:seismicbands}, above $\sim1$\,Hz.

\ac{LLO} is located 7\,km from the town of Livingston, and only 3\,km from a railway line used daily by cargo trains \cite{Daw:2004qd}, plus, the land surrounding \ac{LLO} is used for timber harvesting.
The site is only 130\,km from the Gulf of Mexico, and is subject to violent rain and windstorms.

Both sites are also subject to noise from earthquakes occurring almost anywhere on Earth, and to microseismic noise from oceanic activity due to their relatively short separation from the nearest coastline.
These distant events are sources of noise below $\sim1\,$Hz.

Due to the softer composition of the surrounding geological landscape, seismic noise was worse at \ac{LLO} during early science runs, so the decision was taken to install an active seismic isolation system on L1 before \ac{S4}.
The \ac{HEPI} feed-forward system damps low-frequency noise by using signals from the onsite seismometers to control movement of the vacuum chambers for the end test masses.
This particular system was not installed at \ac{LHO} before \ac{S6} -- although other isolation systems were used -- but will be installed as part of the \ac{aLIGO} project \cite{Abbott:2002cr}.

The \ac{LIGO} instruments are designed to be sensitive in the range $40-7000$\,Hz \cite{Abbott:2007kv}, so one may be forgiven for assuming that seismic noise below $30$\,Hz should not affect sensitivity in the detection band.
However, \gls{upconversion} has been a problem during Initial and Enhanced \ac{LIGO}, caused by a number of factors related to ground motion, for example scattered light \cite{Accadia:2010sl}.
This effect contaminates the sensitive band of the \ac{LIGO} detectors, meaning seismic noise is an even greater problem than it would be otherwise.

\section{Existing veto methods}\label{sec:existing}
\noindent
The \ac{GW} data stream is not the only information drawn from the \ac{LIGO} detectors.
Thousands of auxiliary data channels are recorded, containing control and error signals from instrumental systems, and measurements from the \acp{PEM}.
These data are analysed in order to study and improve detector performance, but also to identify and remove \acp{glitch} that can mimic \acp{GW}.
\Gls{veto} segments can be constructed around excess noise if it is known to couple into the detection channel.

In \ac{S5} and \ac{S6}, \glspl{veto} were constructed by two methods \cite{Blackburn:2008ah, Abadie:2011dc, Slutsky:2010ff, Isogai:2010, Smith:2011an}.

The first method relied on known physical couplings between an auxiliary subsystem and the \ac{GW} data, whereby when a correlation is understood, the time stream of a particular auxiliary channel is analysed, and times for which a certain threshold was exceeded are recorded.
Simple, but highly effective examples include overflows in analog-to-digital converters, and \glspl{light dip}.

The second method replaces known couplings with statistics, applying the \ac{KW} wavelet-based algorithm \cite{Chatterji:2004} to data in auxiliary channels with negligible sensitivity to \acp{GW}, producing lists of \glspl{trigger}.
These events are then tested for time-coincidence with \glspl{trigger} in the \ac{GW} data, indicating whether that candidate \ac{GW} event was likely to be of astrophysical origin.
\Glspl{veto} are constructed around a subset of \glspl{trigger} in the auxiliary data chosen to maximise \gls{efficiency} whilst minimising \gls{deadtime}.

In both methods, highly effective \glspl{veto} are those with a high ratio of \gls{efficiency} to \gls{deadtime}.
Different implementations \cite{Isogai:2010, Smith:2011an} were used in the searches for unmodelled bursts \cite{Abbott:2009zi} and \acp{CBC} \cite{Abbott:2009tt} during \ac{S5}, producing comparable results \cite{Blackburn:2008ah}.

\section{Targeted veto methods}\label{sec:method}
\noindent
The methods described in the previous section are subject to shortcomings when applied to seismometer data.
Simple thresholds have to be placed high enough to catch only the worst noise spikes, and so have low efficiency over weekends and evenings (around lower seismic noise as shown in Figure \ref{fig:seisdiff}).
Similarly, the \ac{KW} algorithm is tuned for high-frequency \acp{GWB}, with limited sensitivity to the low-frequency signature of seismic noise, resulting in low trigger numbers and poor statistical significance of coincidences.

In order to produce effective \glspl{veto}, we have devised a novel method to explicitly identify low-frequency seismic events, and construct  \gls{veto} segments to remove this noise from \ac{GW} searches.
This method uses the \textit{$\mathit{\Omega}$-pipeline} tuned specifically for low-frequency performance to generate \gls{trigger} lists highlighting seismic events, and the low-latency inspiral pipeline \textit{Daily iHope} to generate \glspl{trigger} from \ac{CBC} template matched-filtering.
The two are combined by the \textit{HierarchichalVeto} algorithm into lists of time segments during which seismic noise has polluted the \ac{GW} analysis.

\subsection{The $\mathit{\Omega}$ pipeline}
The $\mathit{\Omega}$\textit{-Pipeline} is a burst detection algorithm developed within \ac{LIGO} as a combination of the \textit{Q Pipeline} \cite{Chatterji:2005t} and \textit{X-Pipeline} \cite{Sutton:2009gi} and used, during \ac{S6}, for low-latency detection of \ac{GW} events to trigger electromagnetic followup \cite{Abadie:2011em}.
The single-detector triggers were also used for \ac{DQ} investigations.

\begin{figure}
  \centering
  \includegraphics[width=0.98\linewidth]{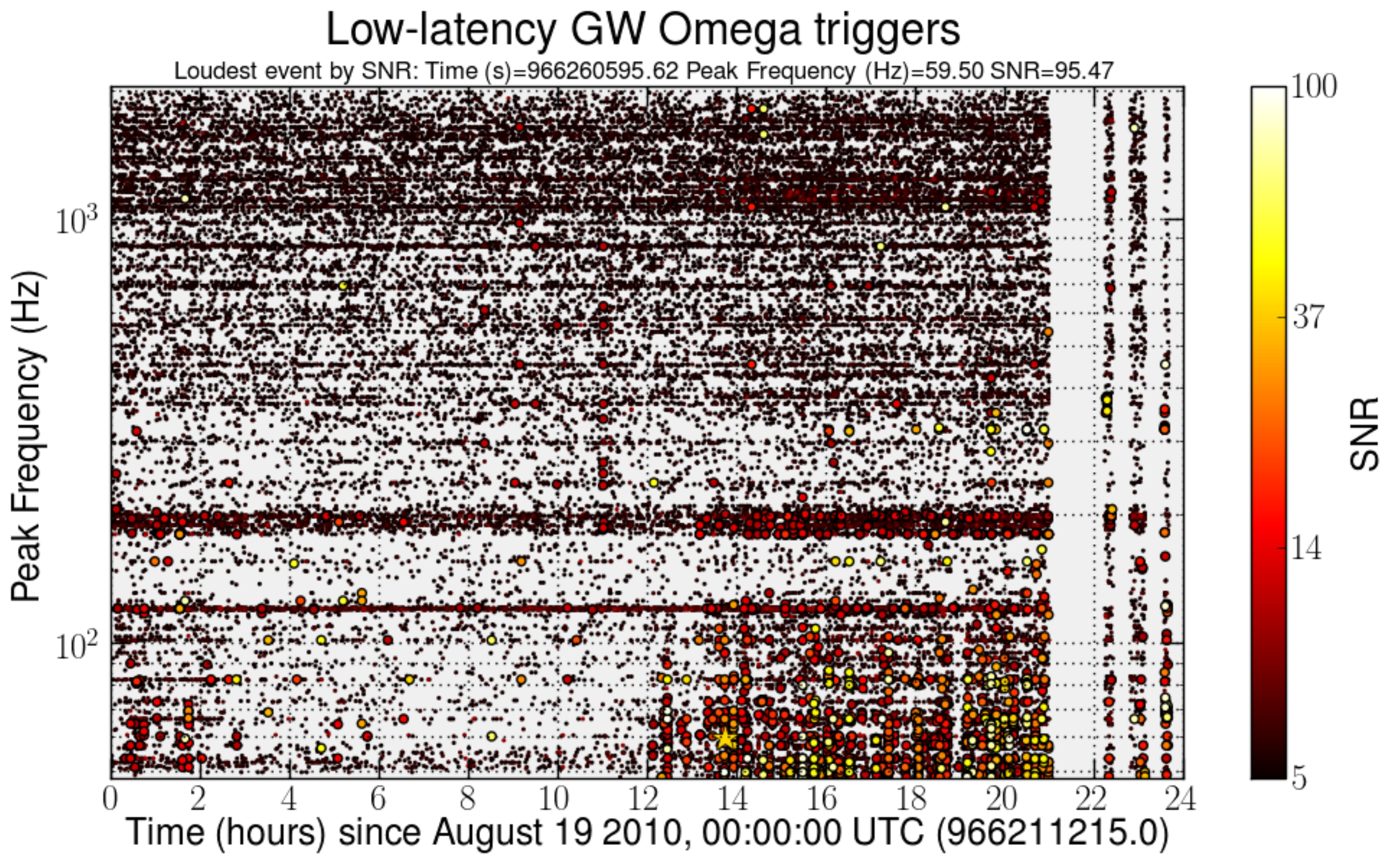}
  \caption{\label{fig:omegagram}
A 24-hour time-frequency map from the $\Omega$-Pipeline of data from the \ac{GW} readout of the H1 detector for the same period as Figure \ref{fig:seisdiff} (top).
12:00 UTC (05:00 local time) onwards are subject to increased noise at the lowest frequencies due to excess seismic noise from the working day.
The stripes with no events at the end of the period indicate that the detector was not operational.}
\end{figure}

The algorithm is based on the Q Transform \cite{Brown:1991} and projects detector data, $s(t)$, onto a bank of windowed complex exponentials of the following form:
\begin{equation}
\label{eq:singaussian}
S(\tau,f,Q) = \int_{-\infty}^{\infty} s(t) w(t-\tau,f,Q)\exp(-i 2 \pi f t)\,dt,
\end{equation}
where $w$ is a time-domain window centred on $\tau$, $f$ is the central frequency, and $Q$ is the quality factor.
An example of the output of the $\Omega$-Pipeline applied to low-latency gravitational wave readout data is show in Figure \ref{fig:omegagram}.
A high density of low \ac{SNR} (black) triggers is expected from Gaussian noise, but the higher \ac{SNR} events (white) indicate increased noise at low frequencies, known to be correlated with the high seismic activity shown in Figure \ref{fig:seisdiff}.

As a result of using both \ac{KW} and $\Omega$-Pipeline for \ac{DQ} studies, direct comparisons were drawn on the performance of each, especially in frequency reconstruction at low frequency.
It was found that, in its standard implementation, the $\Omega$-Pipeline gave much greater low frequency sensitivity, and better frequency resolution.

\subsection{Parameterisation of the $\Omega$-Pipeline for seismic noise}
The low-latency $\Omega$-Pipeline analysis used a parameter set tuned for performance in the detection band, with a frequency range of 48--2048\,Hz, and analyses performed using 64-second-long data segments to estimate the background noise spectrum. 
As can be seen in Figure \ref{fig:omegagram}, the frequency range is such that the seismic band is almost completely ignored.
However, significant \ac{SNR} is recorded up to around $200$\,Hz that can be attributed, by time-coincidence, to seismic noise upconversion.

In order to improve performance when applied to seismometer data, the parameter space was split into two sets: the anthropogenic band, above 2\,Hz, and very low frequency seismic activity, below 2\,Hz.
The following paragraphs detail the changes made to tune the $\Omega$-Pipeline algorithm for each frequency band, describing three key parameters.
The \textit{sampling rate} defines the highest frequency (half the sample rate) of the data to be filtered, and the \textit{frequency range} gives the complete span of frequencies searched.
The \textit{block duration} defines the length of period used to estimate the \ac{PSD} of the detector, set as a power of 2 (for ease of \ac{FFT} computation. In order that a small number of loud events do not affect the measurement of the background, we use a duration significantly longer than the longest resolvable events, and use a median-mean average method to accurately measure the background noise.

\subsubsection{The anthropogenic band, $\mathit{>2}$\,Hz}
\mbox{}\\
As described in Section \ref{sec:seismic} the seismic band extends upwards in frequency to a few tens of Hz.
Short, high-frequency events may corrupt the calculation of the background around them for a longer time, shadowing lower-frequency, lower-amplitude events. Lowering the sampling frequency to $64$\,Hz\footnote{The $\Omega$-Pipeline search for \acp{GW} downsamples the readout data to 4096\,Hz, while the seismometers are only sampled at 256\,Hz} filtered out any high frequency seismic noise, allowing longer time-scale events to be triggered by the search.
The power spectrum was drawn from blocks of 4096\,s, meaning a number of discrete seismic events above 2\,Hz could be individually resolved above the background.

\subsubsection{The earthquake band, $\mathit{<2}$\,Hz}
\mbox{}\\
This band was chosen specifically to target long-distance earthquakes, that, as described in Table \ref{table:seismicbands}, add noise down to $0.01$\,Hz for up to several hours.
Here the sampling frequency could be reduced to 4\,Hz, eliminating higher frequency disturbances, with a minimum frequency of $0.01$\,Hz, while blocks of 65536\,seconds were used to allow accurate \ac{PSD} estimation in the presence of hour-long earthquake events.\\
\begin{table}
  \begin{center}
    \begin{tabular}{l||c||c|c}
      \hline
      \hline
      Parameter & Untuned Value & \multicolumn{2}{c}{Tuned value}\\
      &&$<2$\,Hz&$>2$\,Hz\\
      \hline
      \hline
      Sample frequency & 4096\,Hz & 4\,Hz & 64\,Hz\\
      Frequency range & $48-2048$\,Hz & $0.01-2$\,Hz & $2-32$\,Hz\\
      Block duration & $64$\,s & $65536$\,s & $4096$\,s \\
    \end{tabular}
  \end{center}
  \caption{The parameter sets applied to the $\Omega$-Pipeline search algorithm before and after tuning for low-frequency performance.}
\end{table}

\begin{figure}
  \centering
  \subfloat{
    \includegraphics[width=0.48\linewidth]{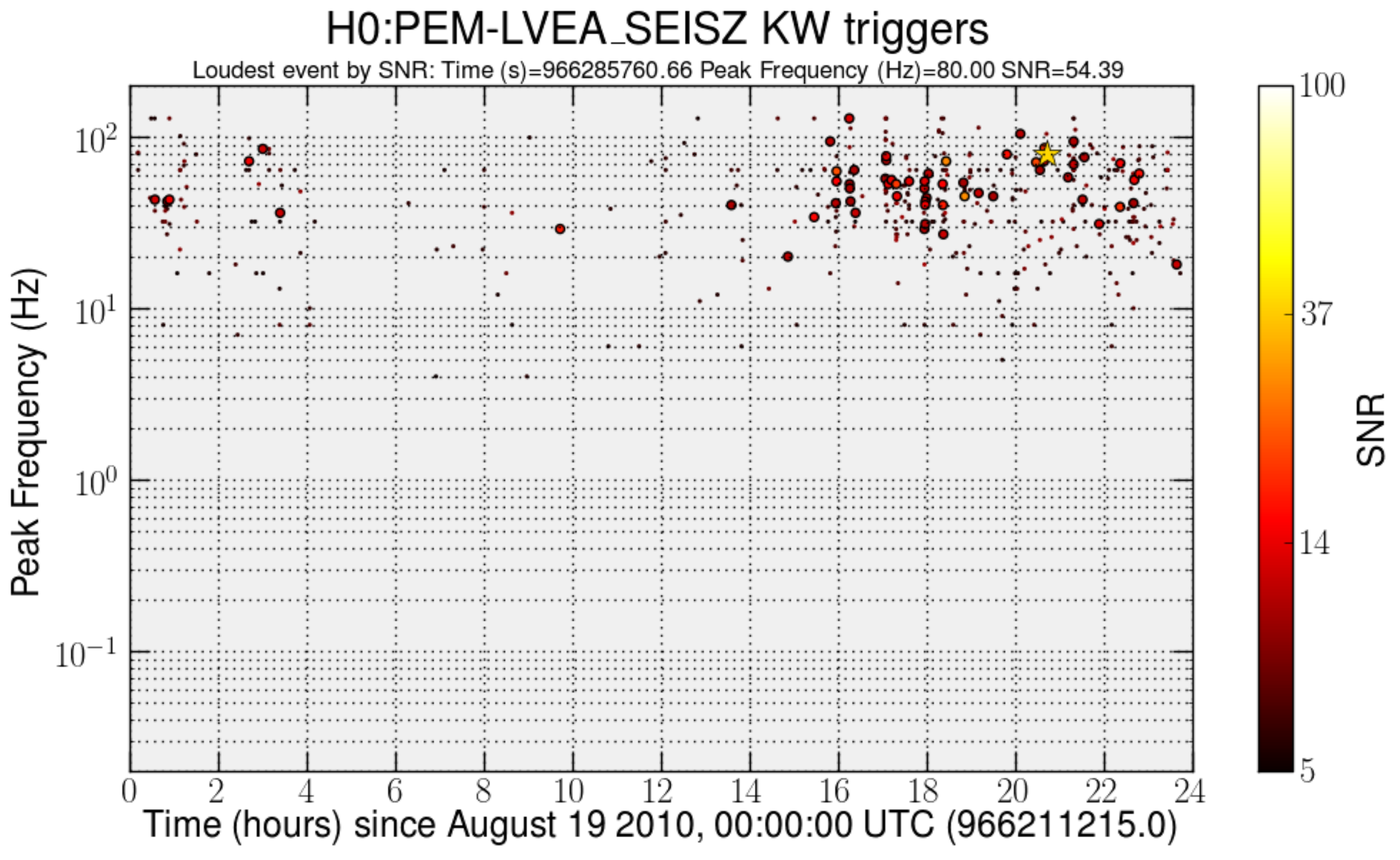}\label{subfig:kwseisgram}}
  \subfloat{
    \includegraphics[width=0.48\linewidth]{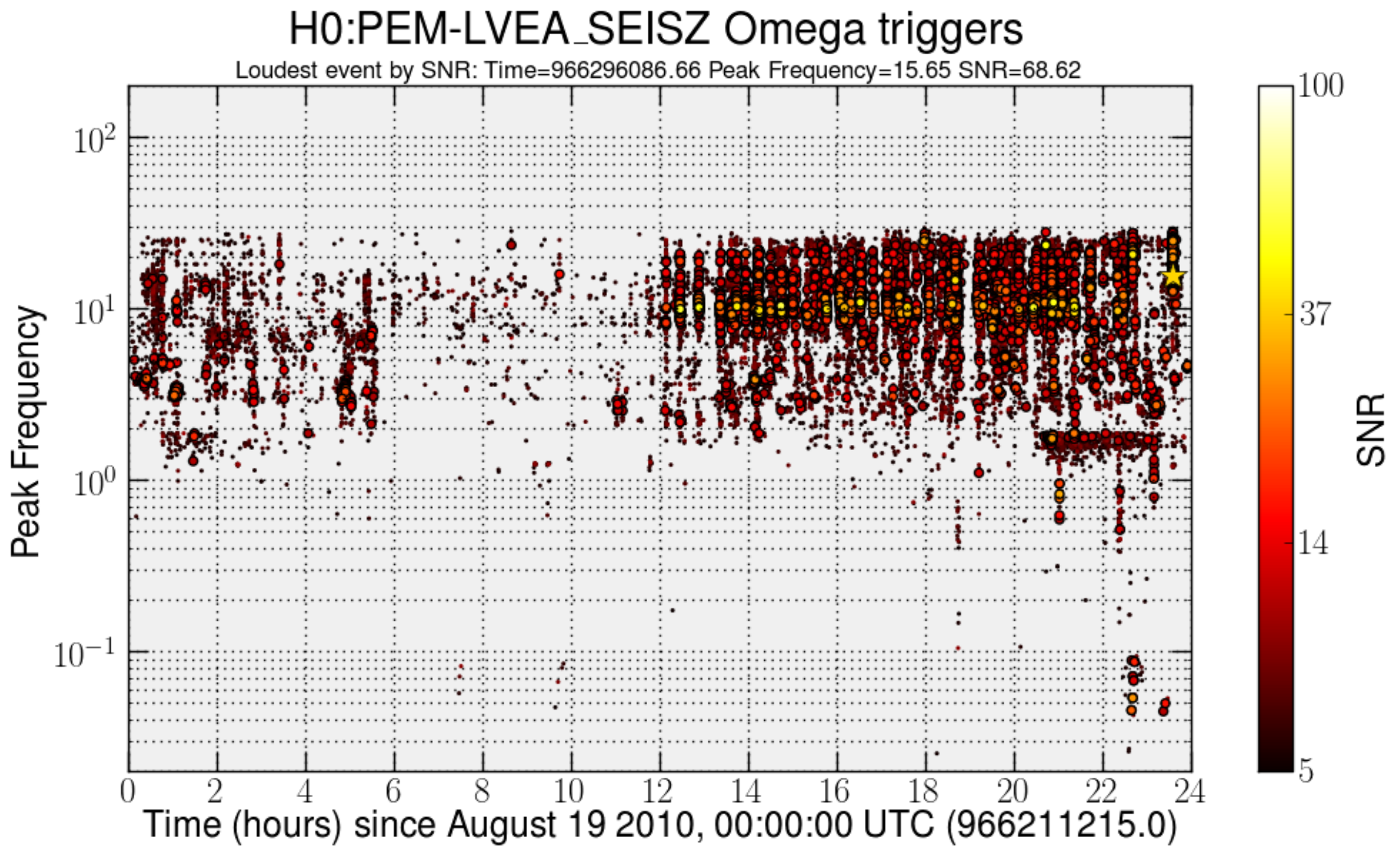}\label{subfig:omegaseisgram}}
  \caption{\label{fig:seisgrams}
Examples of the untuned \ac{KW} and tuned $\Omega$-Pipeline algorithms applied to seismometer data.
The left figure, \ref{subfig:kwseisgram}, shows the lack of sensitivity in the untuned \ac{KW} analysis, while that on the right, \ref{subfig:omegaseisgram}, has many orders of magnitude more events.
Comparing to Figure \ref{fig:omegagram} we can see loud triggers around 10\,Hz after 12:00 UTC, but also triggers with \ac{SNR} above 10 below 0.1\,Hz.}
\end{figure}

\noindent This method was applied to the four main seismometers at \ac{LHO}: EX, EY, LVEA and VAULT\footnote{The EX and EY seismometers sit outside of the vacuum chambers containing the end test masses for the X- and Y-arms respectively, the \ac{LVEA} seismometer sits beside the chamber housing the \ac{GW} readout photodetector, and the VAULT seismometer is in an underground chamber a small distance from the \ac{LVEA}.}; and the three at \ac{LLO}: EX, EY and LVEA (LLO has no VAULT seismometer).
As can be seen in Figure \ref{subfig:omegaseisgram}, the new parameter sets allows a huge increase in the number of triggers produced by the $\Omega$-Pipeline.
The density of triggers has be greatly increased, especially around noisier times, with events recorded with frequencies as low as 0.03\,Hz.

\subsection{Low-latency inspiral triggers -- \textit{Daily iHope}}
The joint \ac{LIGO}-Virgo \ac{CBC} group uses the \textit{iHope} pipeline to search for \acp{GW} produced by binary coalescences.
It is described more fully in \cite{Abbott:2009tt, Abbott:2009qj}.
Seismic noise has been known to contribute significantly to the noise background esimates in these searches, and so creating good vetoes specifically for \ac{CBC} searches was a major goal of this work.
Here we summarize the key points and discuss the changes made for daily running of iHope in order to provide triggers to analyse alongside the $\Omega$-Pipeline triggers from seismic data.

\begin{figure}
  \includegraphics[width=0.98\linewidth]{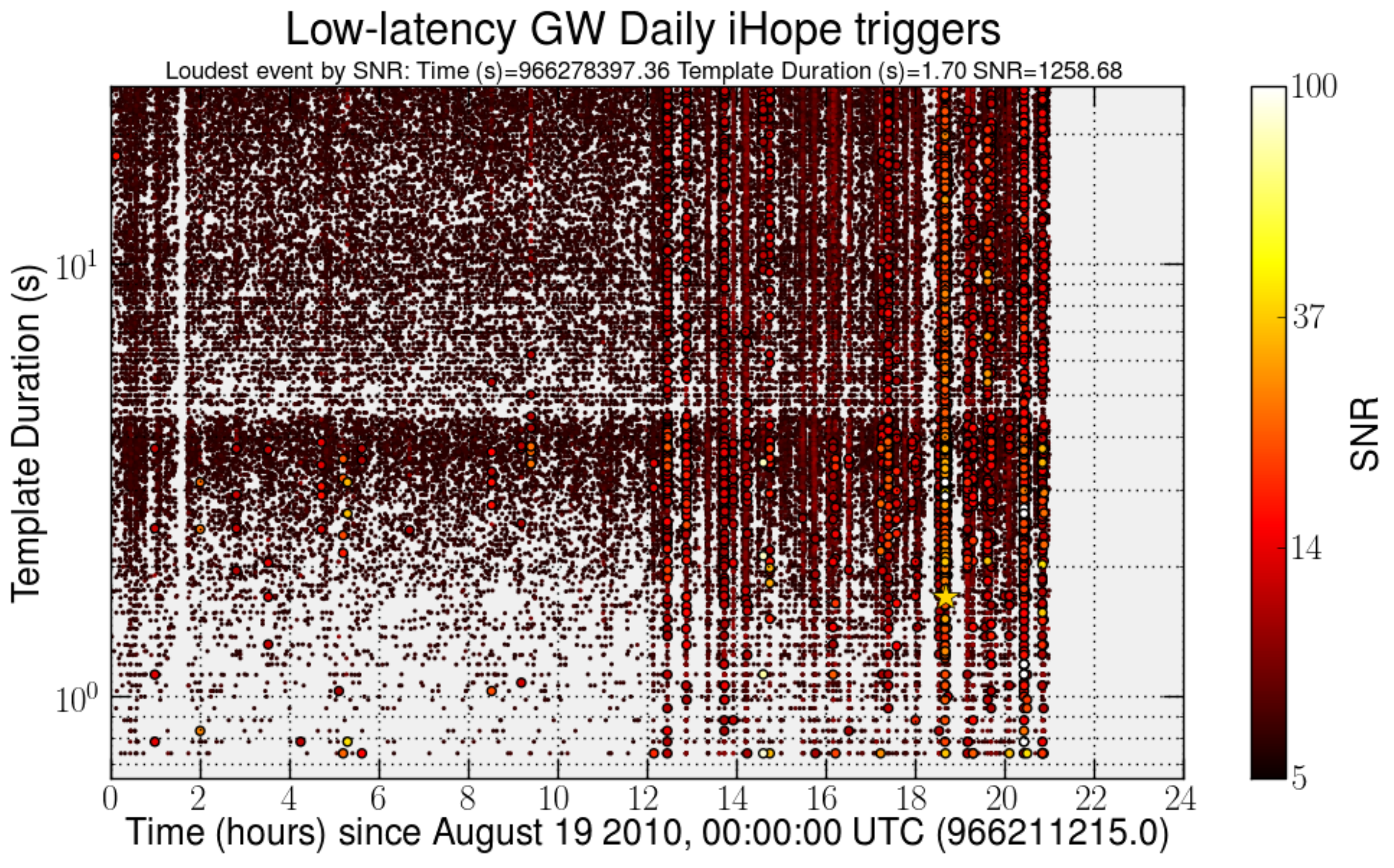}
  \caption{A 24-hour template duration versus time map of triggers from the Daily iHope pipeline.
\ac{CBC} templates are better characterised by a duration rather than a frequency, as they sweep through a range of frequencies.
Comparing to Figures \ref{fig:seisdiff} and \ref{fig:omegagram}, we see the same excess of triggers after 12:00 UTC, but here the excess noise at low frequency results in triggers across the entire template bank.
No data is analyzed after 21:00 UTC because ihope requires at least 2048 contiguous seconds to estimate the \ac{PSD} and all data after this time was in smaller segments.}
  \label{fig:daily_ihope_glitchgram}
\end{figure}

Daily iHope is a templated, matched-filter search using restricted, stationary phase, frequency-domain waveforms of the form
\begin{equation}
  \label{eq:spa_template}
  \tilde{h}(f; M, \eta) 
  =
  \frac{2GM_\odot}{c^2 r} 
  \left( \frac{5 M \eta}{96 M_\odot} \right)^\frac{1}{2}
  \left( \frac{M}{\pi^2 M_\odot} \right)^{\frac{1}{3}}
  f^{-\frac{7}{6}}
  \left( \frac{G M_\odot}{c^3} \right)^{-\frac{1}{6}}
  e^{ i \Psi(f; M, \eta) },
\end{equation}
where $M=m_1+m_2$ is the total mass of the binary, $\eta = m_1m_2/M^2$ is the symmetric mass ratio.
A static bank of such templates spanning the total mass range from $2\,M_\odot - 25\,M_\odot$ was used for each interferometer, based on the layout at a quiet time in each instrument, with a minimal match of 0.95 for the region above a chirp mass ($M\eta^{3/5}$) of 3.46\,$M_\odot$  and 0.5 below that.
This distribution would not be good enough for an astrophysical search, but was shown to be adequate for identifying short-duration glitches that match the higher-mass (shorter) templates better.
This allowed for short-duration glitches, corresponding to higher-mass inspiral templates, to be flagged with large \ac{SNR}.
An example of the output of Daily iHope is shown in Figure \ref{fig:daily_ihope_glitchgram}.

\subsection{Veto generation -- HierarchichalVeto}
The seismic triggers from the $\Omega$-Pipeline, and the \ac{CBC} triggers from Daily iHope were used to idenfity times of seismic noise using the statistical algorithm \ac{HVeto} \cite{Smith:2011an}.

The \ac{HVeto} algorithm tests the statistical significance of time-coincidence between triggers from one channel, nominally the \ac{GW} data channel, and those from auxiliary channels. The significance statistic is defined as
\begin{equation}
\label{eq:hvetosig}
S_n(x) = -\log_{10}\left(\sum_x^{\infty} P_{poi}(\mu,x^{\prime})\right),
\end{equation}
where $x^{\prime}$ is the number of coincident events, $\mu$ is the expected number of random coincidences given the trigger rates in the two channels, $x$ is the series of non-negative integers, and $P_{poi}$ is the Poisson probability distribution function. The significance is calculated for all channels in a two-dimensional space of time-coincidence window, $T_{\mbox{\footnotesize win}}$\footnote{Low-frequency events have a long duration whose maximum coupling time is not known. Also, many short-duration \acp{glitch} in an auxiliary system can take a certain time to couple into the \ac{GW} output.}, and \ac{SNR} threshold, $\rho_c$.

The most significant point on the $(T_{\mbox{\footnotesize win}},\rho_c)$ plane is chosen for each auxiliary channel, with the loudest channel by significance selected. Veto times are constructed by generating segments of width $T_{\mbox{\footnotesize win}}$ around all triggers with \ac{SNR} above $\rho_c$ in that auxiliary channel. These segments are then removed from the analysis -- allowing the next round to be `won' by a (generally) different auxiliary channel containing less significant coincidences -- and the procedure repeated until the significance of the loudest channel does not exceed a given stopping point. In this way, vetoes are generated hierarchichally, allowing for little redundancy between different channels.

Several modifications were made to this algorithm in order to test and run on the new seismic triggers. Testing was completed in order to construct a new $(T_{\mbox{\footnotesize win}},\rho_c)$ plane relevant for the long-duration events from the seismic data, spanning $0.1<=T_{\mbox{\footnotesize win}}<10$\,s and SNRs 10--300. Alongside this, as described in the caption to Figure \ref{fig:daily_ihope_glitchgram}, modifications were made to first read and understand the new Daily iHope triggers, and use the relevant new parameters.

\section{Results}\label{sec:results}
\noindent
The method described above was used in the construction of the \ac{LIGO} seismic veto  dubbed \glsentryname{SeisVeto} for \ac{S6}.
Here we present the results for a test sample of data, spanning June 26 -- August 6 2010.
The results for \ac{LHO} are shown in Table \ref{table:h1results} and those for \ac{LLO} in Table \ref{table:l1results} with each row giving the statistics for the most significant channel in each frequency band, in addition to the cumulative results for the entire period\footnote{The cumulative results include all rounds passing selection criteria in each band, not just the most significant channel.}.
\begin{table}  \centering
  \subfloat[Results for H1.\label{table:h1results}]{
    \begin{tabular}{c|c|c|c|c}
      \hline\hline
      Freq.
Band (Hz) & Loudest Channel & Significance & Efficiency (\%) & Deadtime (\%)\\
      \hline\hline
      0-1 & EX & 1455.21 & 3.26 & 0.15\\
      1-3 & EY & 355.37 & 3.19 & 0.71\\
      3-10 & LVEA & 12024.98 & 22.11 & 1.24\\
      10-32 & LVEA & 41042.78 & 35.99 & 1.04\\
      \hline\hline
      \multicolumn{3}{r|}{Cumulative, all rounds} & 62.44 & 5.94\\
      \hline\hline
    \end{tabular}}\\
  \subfloat[Results for L1.\label{table:l1results}]{
    \begin{tabular}{c|c|c|c|c}
      \hline\hline
      Freq.
Band (Hz) & Loudest Channel & Significance & Efficiency (\%) & Deadtime (\%)\\
      \hline\hline
      0-1 & \multicolumn{2}{c|}{\footnotesize No channels passing selection criteria} &  0 & 0 \\
      1-3 & LVEA & 960.13 & 1.51 & 0.06\\
      3-10 & LVEA & 420.55 & 0.88 & 0.06\\
      10-32 & EX & 1601.22 & 2.29 & 0.07\\
      \hline
      \hline
      \multicolumn{3}{r|}{Cumulative, all rounds} & 6.95 & 0.60 \\
      \hline\hline
    \end{tabular}}
  \caption{\label{table:results}
\ac{HVeto} results from coincidence between new $\Omega$-Pipeline triggers from seismometer data and low-latency inspiral triggers for the \ac{LIGO} detectors operating in \ac{S6}.
Shown are the most significant (loudest) channels for each frequency band, and the cumulative statistics for the entire analysis.}
\end{table}

For H1, the rounds contribute to give a cumulative efficiency of $62.5$\,\%, with a cumulative deadtime of $5.94$\,\%.
This means that almost two thirds of all triggers produced by the low-latency inspiral pipeline are occurring in a small amount of time, which is coincident with high seismic noise.
This statistic alone outlines the problem caused by seismic noise.

It should not be surprising that the most significant channel for the two higher frequency bands should be the \ac{LVEA} seismometer.
This building is closer than any other to a major road, so experiences the highest magnitude of seismic noise from traffic and close anthropogenic noise, especially trucks serving the \ac{USDOE} Hanford site, and also houses the majority of interferometer control optics and subsystems, notably the \ac{GW} readout photodetector.

For L1 we can see much lower statistical significance of the correlation between seismic noise and the readout signal.
This can be attributed in part to the improvements from the HEPI feedforward system for the Livingston instrument, but also to the different nature of the seismic environment relative to \ac{LHO}.
However, despite a lower efficiency, the ratio of efficiency to deadtime is still above 10, highlighting the statistical correlation between seismic noise and low-latency \ac{CBC} \glspl{trigger}.

\section{Summary}\label{sec:summary}
\noindent 
In this paper, we have highlighted the problem of transient seismic noise in \ac{GW} detection, and presented a new method to not only identify, but remove, times of high noise from short-duration \ac{GW} searches.
We have demonstrated a highly effective \gls{veto}, with large \gls{efficiency}-to-\gls{deadtime} ratio, that has been crucial in removing the worst of the transient detector noise whilst leaving as much searchable time as possible.
This method was applied to the searches for \ac{GWB} and \ac{CBC} signals in the final part of \ac{S6}, and was seen to have a dramatic effect on the background, \cite{Mciver:2011am}.

\acf{aLIGO} is a major upgrade program that will see the sensitive distance of the \ac{LIGO} detectors increase by a factor of 10, giving a factor of 1000 in sensitive volume.
This should mean regular detections of \ac{GW} transients from \ac{CBC} events \cite{Abadie:2010cf}.
However, it is likely that there will still be non-stationarities in the data from seismic events, and other sources.
The method introduced here will allow us to remove them, increasing search sensitivity, and also gives a highly tuned means of directing site scientists to coupling noise sources in a newly commissioned machine.

Seismic noise was chosen as an obvious starting point, given the prevalence of glitches of seismic origin, and a prior lack of an effective veto method.
This method can be generalised to any and all susbsystems of the next generation of interferometers by tuning a \ac{GW} burst detection algorithm on the appropriate data channels and has the potential to lead to a great increase in search sensitivity as a result of the above benefits.

\section*{Acknowledgements}\label{sec:acknowledgements}
\noindent
The authors would like to thank Gabriela Gonzalez, Jessica McIver, Greg Mendell, Laura Nuttall, and all members of the Detector Characterization group, the $\Omega$-Pipeline team, and the HierarchichalVeto team for discussions. DMM was supported by a studentship from the Science and Technology Facilities Council. SF was supported by the Royal Society. BH was supported by NSF grant PHY-0970074 and the UWM Research Growth Initiative. LP and AL were supported by NSF grant PHY-0847611. JRS was supported by NSF grants PHY-0854812 (Syracuse) and PHY-0970147 (Fullerton).

\newpage

\bibliographystyle{unsrt}
\bibliography{references.bib}
\end{document}